\documentclass{ws-ijmpa}

\usepackage{amsmath}
\usepackage{amssymb}
\def\be{\begin{equation}}
\def\ee{\end{equation}}
\def\ba{\begin{array}}
\def\ea{\end{array}}
\def\beqn{\begin{eqnarray}}
\def\eeqn{\end{eqnarray}}

\def\bt{\begin{tabular}}
\def\et{\end{tabular}}
\def\bc{\begin{center}}
\def\ec{\end{center}}

\begin{document}
\markboth{Samandeep Sharma, Priyanka Fakay, Gulsheen Ahuja,
Manmohan Gupta} {Clues towards unified textures}

%
\catchline{}{}{}{}{}
%

\title{\bf{Clues towards unified textures}}

\author{SAMANDEEP SHARMA$^*$}


\author{PRIYANKA FAKAY$^*$}

\author{GULSHEEN AHUJA$^*$}


\author{MANMOHAN GUPTA$^* \dagger$}
\address{\it {$^*$ Department of Physics, Centre of Advanced Study, P.U.,
 Chandigarh, India.\\
 $\dagger$ mmgupta@pu.ac.in}}

\maketitle

\begin{history}
\received{Day Month Year}
\revised{Day Month Year}
\end{history}

\begin{abstract}
The issue of texture specific mass matrices has been discussed by
incorporating Weak Basis transformations and the concept of
`naturalness'. Interestingly, we find that starting from the most
general mass matrices, one can arrive at texture four zero mass
matrices which can fit both quark as well as lepton mixing data
and are similar to the original Fritzsch ansatze.

\end{abstract}

\ccode{PACS numbers:12.15.Ff,14.60.Pq}


\renewcommand{\baselinestretch}{1.50}\normalsize
\section{Introduction}
 Fermion masses and mixings constitutes one of the most important aspects of
 Flavor Physics. Understanding the vast spectrum of fermion masses, spanning many orders of
 magnitude,
  in a unified framework is one of the biggest challenges. For a better appreciation of
  the problem, it is desirable to consider the present spectrum of fermion
  masses, e.g.,
 \bc
 $m_u , m_d\sim 10^{-3}GeV,~ m_s\sim 10^{-1} GeV,~~~ m_c ,m_b\sim
1GeV,~~ m_t\sim 10^{2}GeV$ ,\\$ m_e\sim 10^{-3} GeV,~~ m_{\mu}\sim
10^{-1} GeV, ~~ m_{\tau}\sim 1GeV$ ,~~\\$ m_{\nu} \lesssim
10^{-11} GeV.$
\ec
The above ranges of masses clearly span over 13 orders of
magnitude. In case the theory requires the existence of right
handed neutrinos, with the mass range ~~$\sim 10^{9}- 10^{16}$ GeV,
the fermion masses would then cover almost 30 orders of
magnitude.
\par It is well known that the quark mixing angles are  hierarchical, e.g. $s_{12}\sim
0.22$, $s_{23}\sim 0.04$, $s_{13}\sim 0.004$, whereas the lepton
mixing angles do not show any hierarchy.  The non zero value of
$\theta_{13}$\cite{theta13-1}\cdash\cite{theta13-4}, on the one
hand, restores the parallelism between the mixings of quarks and
leptons, on the other hand, its unexpectedly `large' value
signifies the differences between these, the leptonic mixing
angles being large as compared to the quark counterparts values.
Further, the issue of neutrinos being Dirac like or Majorana
particles is still an open question for  physicists since  Dirac
neutrinos have not yet been ruled by experimental data. The
problem assumes several dimensions when one finds that in the case
of quarks not only the mixing angles but the quark masses also
show distinct hierarchy, this being in sharp contrast to the case
of neutrinos wherein neither the mixing angles nor the masses show
any distinct hierarchy. Since the mixing angles are related to the
corresponding mass matrices therefore formulating viable fermion
mass matrices becomes all the more complicated especially when
quarks and leptons have to be described in a unified framework.

\par The theoretical understanding of fermion masses and mixings
   proceeds along two approaches, i.e.,  `top-down' and `bottom-up'.  Despite
   large number of attempts from the top-down
perspective, such as grand
unification\cite{gut-1}\cdash\cite{gut-6},
supersymmetry\cite{susy-1}\cdash\cite{susy-3},
compositeness\cite{comp-1}\cdash\cite{comp-5},
superstrings\cite{string-1}\cdash\cite{string-3}, etc., we are not
in a position to have a compelling theory of flavor dynamics.
Bottom-up approach consists of finding the phenomenological
fermion mass matrices
       which are in tune with the low energy data, i.e., observables
        like quark and lepton
       masses, mixing angles in both the sectors, angles of the unitarity triangle in the quark sector, etc..

\par Texture specific mass matrices provide a very good example of
bottom-up approach.  For a detailed and comprehensive review we
refer the reader to Fritzsch and Xing\cite{fritzschreview}, along
with a very recent one by Gupta and Ahuja\cite{singreview}. A
viable formulation of mass matrices which incorporates the low
energy data pertaining to quark sector as well as lepton sector is
very desirable. It becomes particularly desirable in case  we want
to find viable, finite set of texture specific mass matrices,
which then can provide vital clues for their formulation within
the top-down approach.

\par Several attempts have been made to describe fermion mass matrices in a unified
  manner\cite{chen,joshipura} within the framework of grand unified theories (GUTs). Efforts
  have also been made to reconstruct viable fermion mass matrices keeping in mind the
  low energy data as well as to integrate textures within the GUTs\cite{singreview,talk}. However,
  the issue of finding whether a viable set of mass matrices which can accommodate the quark mixing as well as lepton mixing
  data within the texture framework has not been addressed yet. The purpose of present work, therefore,
  is to find a minimal viable set of texture specific mass
 matrices which are in agreement with Weak Basis (WB) transformations as well as the criterion of natural
 mass matrices. To this end, we first explore the possibility of finding texture specific quark mass matrices
 starting from the most general mass matrices. The issue of compatibility of such mass matrices with
 the lepton mixing data has also been discussed briefly.

\par The detailed plan of paper is as follows. To make the manuscript self contained, in section \ref{wbt}
we discuss briefly the essentials of Weak Basis transformations and in section \ref{natural} we encapsulate the idea of
naturalness. In section \ref{section1}, we present the analysis pertaining to
the general texture specific  quark mass matrices within
the framework of Standard Model (SM) incorporating Weak Basis transformations and naturalness. Without
going into details, in section \ref{t2l}
we present the status of the lepton mass matrices for the same. Finally, section \ref{con}
summarizes our conclusions.
\section{Weak Basis Transformations\label{wbt}}
 Within the framework of the SM, the quark mass matrices can be considered hermitian without loss of generality,
  encoding all the information about the quark masses and mixings.
  These matrices have a total of
18 real free parameters,  large in number compared to only ten
physical  observables corresponding to six quark masses and four
physical parameters of the Cabibbo-Kobayashi-Maskawa (CKM) matrix.
It should be noted that in the SM one has the freedom to make a
unitary transformation, e.g., $ q_L\rightarrow W q_L~,~ q_R
\rightarrow W q_R $,~~ $ q_L^{\prime}\rightarrow W q_L
^{\prime}~,~ q_R^{\prime} \rightarrow W q_R ^\prime $ under which
the gauge currents  remain real and diagonal but the mass matrices
transform as  \be  M_U \rightarrow W^{\dagger} M_U W~,~M_D
\rightarrow W^{\dagger} M_D W ,\ee
  such transformations
are referred to as `Weak Basis (WB) Transformations'. It can be
easily checked that such transformations preserve the hermiticity
of the mass matrices. It needs to be mentioned that the CKM matrix is
independent of WB transformations, e.g., noting that ($U_U$,
$U_D$) and ($U_U ^{\prime}$, $U_D^{\prime}$) are the respective
diagonalizing
 transformations of $(M_U, M_D)$and $(M^{\prime}_U,
M^{\prime}_D)$ for either of $M_U$ and $M_D$  one  can obtain
\be
 U^{\prime}_{U}=
W^{\dagger} U_U;~~ U^{\prime}_{D}= W^{\dagger} U_D. \ee Using this
result, the mixing matrix for the WB transformed quark mass
matrices can be given as
\be
V_{ckm}^{\prime}=U_u^{\prime \dagger}U_d^{\prime}= (W^{\dagger}
U_u)^{\dagger}(W^{\dagger} U_d)= (U_u)^{\dagger} W
W^{\dagger}U_d=(U_u)^{\dagger}U_d=V_{ckm}.
 \label{eqckm}
 \ee

\par WB transformations have been used\cite{branco-1}\cdash\cite{frxing-2} to obtain
texture specific mass matrices which lead to reduction in the
number of parameters defining the mass matrices without loss of
generality. It can be shown that the general mass matrices can be
reduced to two different types of texture specific up and down
mass matrices e.g., Branco {\it{et al.}}\cite{branco-1,branco-2}
use the WB transformations to obtain the following texture
specific mass matrices \be M_q=\left( \ba {ccc} 0 & * & 0 \\
                     {*} & * & * \\
                     0 & * & * \ea \right),~M_{q^{'}}=\left( \ba {ccc} 0 & * & * \\
                     {*} & * & * \\
                     {*} & * & * \ea \right), ~~~~~~q,~q^{'}=U,D.
                   \ee
In the second possibility, as observed by Fritzsch and
Xing\cite{frxing-1,frxing-2}, one ends up with the possibility
\be
M_q=\left( \ba {ccc} {*} & * & 0 \\
                     {*} & {*} & {*} \\
                     0 & * & * \ea \right), ~~~~~~q=U,D.
                   \ee
Needless to emphasize that the two type of mass matrices are
equivalent. However, we adopt the possibility given by Fritzsch
and Xing as it corresponds to parallel structure for $(M_U, M_D)$
in consonance with most of the GUT models.
\section{Natural mass matrices \label{natural}}
The matrices obtained through WB transformations have large
parametric space due to the extra number of parameters as well as
no restriction on the elements of the mass matrices. In case such
matrices are used to fit the data, it is immediately clear that
one would obtain large number of viable possibilities of mass
matrices, making it difficult to develop models corresponding to
these. In this context, Peccei and Wang\cite{nmm}, in order to
avoid fine tuning, have translated the hierarchy of the quark
mixing matrix to the formulation of `natural mass matrices' which
considerably restricts the parameter space available to the
elements of the mass matrices. It is interesting to note that the
idea of natural mass matrices coupled with the WB transformations
leads to constraints on the elements of the mass matrices. In this
context, we first consider the CKM matrix in the Wolfenstein
parametrization\cite{wolf},

 \be V_{\rm CKM} =
 \left( \ba {ccc} 1-\frac{1}{2} \lambda ^2 & \lambda  &
  A \lambda ^3 (\rho- i\eta )\\
  -\lambda &
 1-\frac{1}{2} \lambda ^2
  & A \lambda ^2 \\
  A \lambda ^3 (1- \rho- i\eta )&
  -  A \lambda ^2 &
 1 \ea \right),  \label{2ckm}  \ee
where $\lambda \sim 0.22,~A \sim 0.82$, $\rho \sim 0.131$ and
$\eta \sim 0.345 $. The above matrix can approximately be written
as \be V_{\rm CKM} =\left( \ba {ccc} 1 & \lambda & 0.3 \lambda^{3}
\\
 -\lambda  & 1 & 0.8\lambda^{2} \\
 0.6\lambda^{3} & -0.8\lambda^{2} & 1
  \ea \right).
\label{ck}
  \ee
To translate the constraints of the above matrix on the mass matrices, we consider a basis in
which either of the two mass matrices $M_U$ and $M_D$ is diagonal. For example, in case we choose $M_D$ to be diagonal
and
 \be
M_{U}=\left( \ba {ccc} a & b & c \\ d  & e & f \\ g & h & i \ea
\right), \label{mu} \ee keeping in mind the relation
$M_U=V_{ckm}^\dagger M_U^{diag} V_{ckm}$ and using eqns.
(\ref{ck}) and  (\ref{mu}), we get
\be
\left( \ba {ccc} a & b & c \\ d  & e & f \\ g & h & i \ea \right)=
\left( \ba {ccc} 1 & \lambda & 0.3 \lambda^{3} \\
 -\lambda  & 1 & 0.8\lambda^{2} \\
 0.6\lambda^{3} & -0.8\lambda^{2} & 1 \ea \right)^{\dagger} \left(
\ba {ccc} m_{u}  & 0 & 0 \\ 0 & m_{c} & 0 \\ 0 & 0 & m_{t} \ea
\right) \left( \ba {ccc}1 & \lambda & 0.3 \lambda^{3} \\
 -\lambda  & 1 & 0.8\lambda^{2} \\
 0.6\lambda^{3} & -0.8\lambda^{2} & 1 \ea \right).
 \ee
Since $ m_{u}  \ll  m_{c} \ll  m_{t} $,  one obtains
\be
\left( \ba {ccc} a & b & c \\ d  & e & f \\ g & h & i \ea
\right)\sim \left( \ba {ccc} m_{c} \lambda ^{2} & -m_{c}\lambda
&0.6m_{t}\lambda^{3}\\ - m_{c} \lambda & m_{c}&
-0.8m_{t}\lambda^{3}\\ 0.6 m_{t} \lambda^{3}  &
 -0.8m_{t}\lambda^{3}   &  m_{t} \ea \right).
 \label{wol}
 \ee
It is interesting to note that the elements (1,3) and (3,1) in the above matrix are
larger than the (1,2) and (2,1) elements. However, we still have the freedom of WB
transformations. Noting that WB transformations do not affect the hierarchy of elements of
mass matrices, one can eliminate the (1,3) and (3,1) elements in the above matrix achieving the
form advocated by Fritzsch and Xing. In the matrix so obtained, one can easily see that the
elements of the mass matrices satisfy the natural hierarchy,
\be
(1,i)\lesssim (2,j) \lesssim (3,3);~~~~~~~~~~i=1,~2,~3, ~~j=2,~3.
\label{nathier}
\ee
Further, noting
that the quark mass eigen values are hierarchical, one can always choose a basis where the elements of the
mass matrices satisfy the `natural hierarchy' given in eqn. (\ref{nathier}) and are of the Fritzsch-Xing form,
\be
 M_{i(i=U,D)}=\left( \ba{ccc}
e_{i} & a _{i} & 0      \\ a_{i}^{*} & d_{i} &  b_{i}     \\
 0 &     b_{i}^{*}  &  c_{i} \ea \right).
\label{fxingt2}
\ee
\section {Naturalness and general mass matrices \label{section1}}
To begin with, we start with the mass matrices defined in
eqn.(\ref{fxingt2}) obtained after imposing WB transformations on
the general mass matrices as well as consider these to be natural.
To study the implications of the quark mixing data on these, we
underline some of the important steps used in our analysis. The
elements of the matrices can be defined as $a_{i}
=|a_i|e^{i\alpha_i}$
 and $b_{i} = |b_i|e^{i\beta_i}$, further $\phi_1 =  \alpha_U- \alpha_D$, $\phi_2= \beta_U-
 \beta_D$. The hermitian matrices $M_i$ ($i=U,D$) can be expressed as $ M_i=
P_i^{\dagger} M_i^r P_i \,, $ where $ P_i= {\rm diag}
(e^{-i\alpha_i},\,1,\,e^{i\beta_i})\,$
 and the real matrices
$M_i^r$ are
\be M_i^r = \left( \ba  {ccc} |e_i| & |a_i| & 0\\
|a_i| & d_i & |b_i|\\
0 & |b_i| & c_i \ea \right) \,.
\ee
The matrices $M_i^r$ can be diagonalized by the orthogonal
  transformation, e.g.,
\be M_i^{\rm diag} = O_i^T M_i^{r} O_i \equiv
  O_i^T P_i M_i P_i^{\dagger} O_i
 \,,   \label{o1}\ee
where \be M_i^{\rm diag} = {\rm diag}(m_1,\,-m_2,\,m_3)\,, \ee
 the subscripts 1, 2 and 3 refer respectively to $u,\, c$ and $t$
  for the $U$ sector and $d,\,s$ and $b$ for the $D$ sector. It
  may be noted that the second mass eigen value is chosen with a
  negative sign to facilitate the construction of the
  diagonalizing transformation $O_i$, for the details in this
  regard we refer the reader to our earlier papers [\cite{singreview,diractable}]. The CKM
matrix in terms of the diagonalizing transformations is given by
\be
V_{\rm CKM} = O_{U}^{T} P_{U} P_{D}^{\dagger} O_{D} =
V_{U}^{\dagger} V_{D}\,,  \label{v}
\ee where the unitary matrices
$V_{U}( = P^\dagger_U \, O_U$) and $V_{D}( = P^\dagger_D \, O_D$)
are the diagonalizing matrices for the hermitian matrices $M_U$
and $M_D$ respectively.

\par For ready reference, the input quark masses
  at the $m_Z$  scale\cite{xing2012} used in the analysis are
 \beqn
 m_u = 1.38^{+0.42}_{-0.41}\, {\rm MeV},~~~~~m_d =
2.82^{\pm 0.48}\, {\rm MeV},~~~~ m_s=57^{+18}_{-12}\, {\rm
MeV},~~~~~~~~\nonumber\\ m_c=0.638 ^{+0.043}_{-0.084}\, {\rm
GeV},~~ m_b=2.86 ^{+0.16}_{-0.06}\, {\rm GeV},~~ m_t=172.1 ^{\pm
1.2} \, {\rm GeV}. ~~~\label{qmasses}
\eeqn
The light quark masses $m_u$, $m_d$ and $m_s$ have been
further constrained by using the following mass ratios\cite{leut}
 \be
 m_u / m_d =0.553 \pm 0.043,~~m_s / m_d =18.9 \pm 0.8~.
 \label{ratios}
 \ee
\par For the purpose of our calculations, apart from imposing the
 constraints of naturalness, we further restrict the parameter space by putting the condition that the
 elements of the diagonalizing transformations $O_U$ and $O_D$ remain real. The phases $\phi_{1}$ and $\phi_{2}$
 are given full variation, whereas the free
parameters $d_U$ and $d_D$ have been given wide variation keeping
in mind the above mentioned constraints. While carrying out the
analysis, we first reproduce $V_{us}$, $V_{ub}$ and $V_{cb}$
corresponding respectively to $s_{12}$, $s_{23}$ and $s_{13}$ as
well as CP asymmetry parameter Sin2$\beta$, this essentially
allows us to reconstruct the corresponding CKM matrix.
\begin{figure}[hbt]
 \begin{minipage} {0.45\linewidth} \centering
\includegraphics[width=2.in,angle=-90]{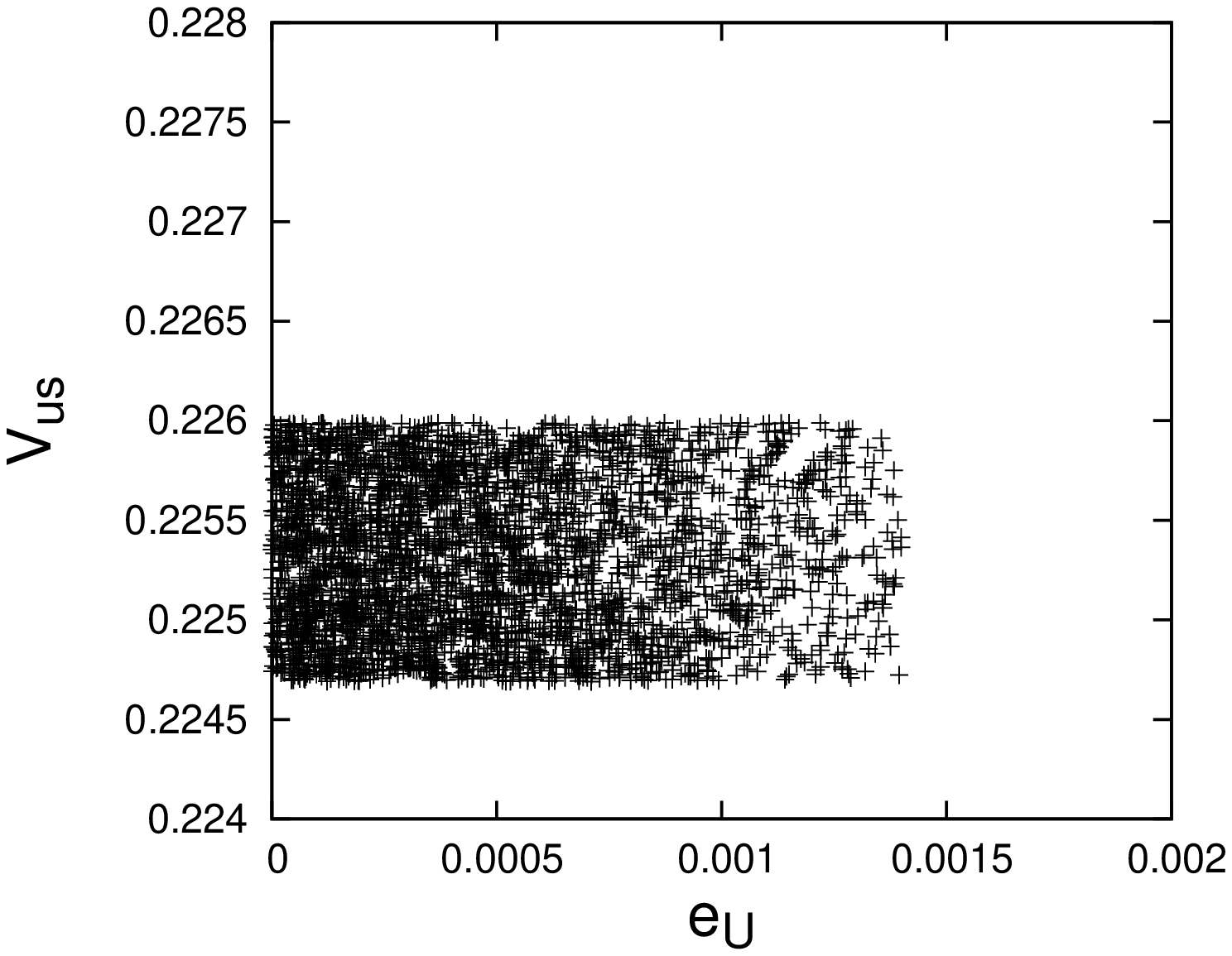}
  \end{minipage}\hspace{0.5cm}
\begin{minipage} {0.45\linewidth} \centering
\includegraphics[width=2.in,angle=-90]{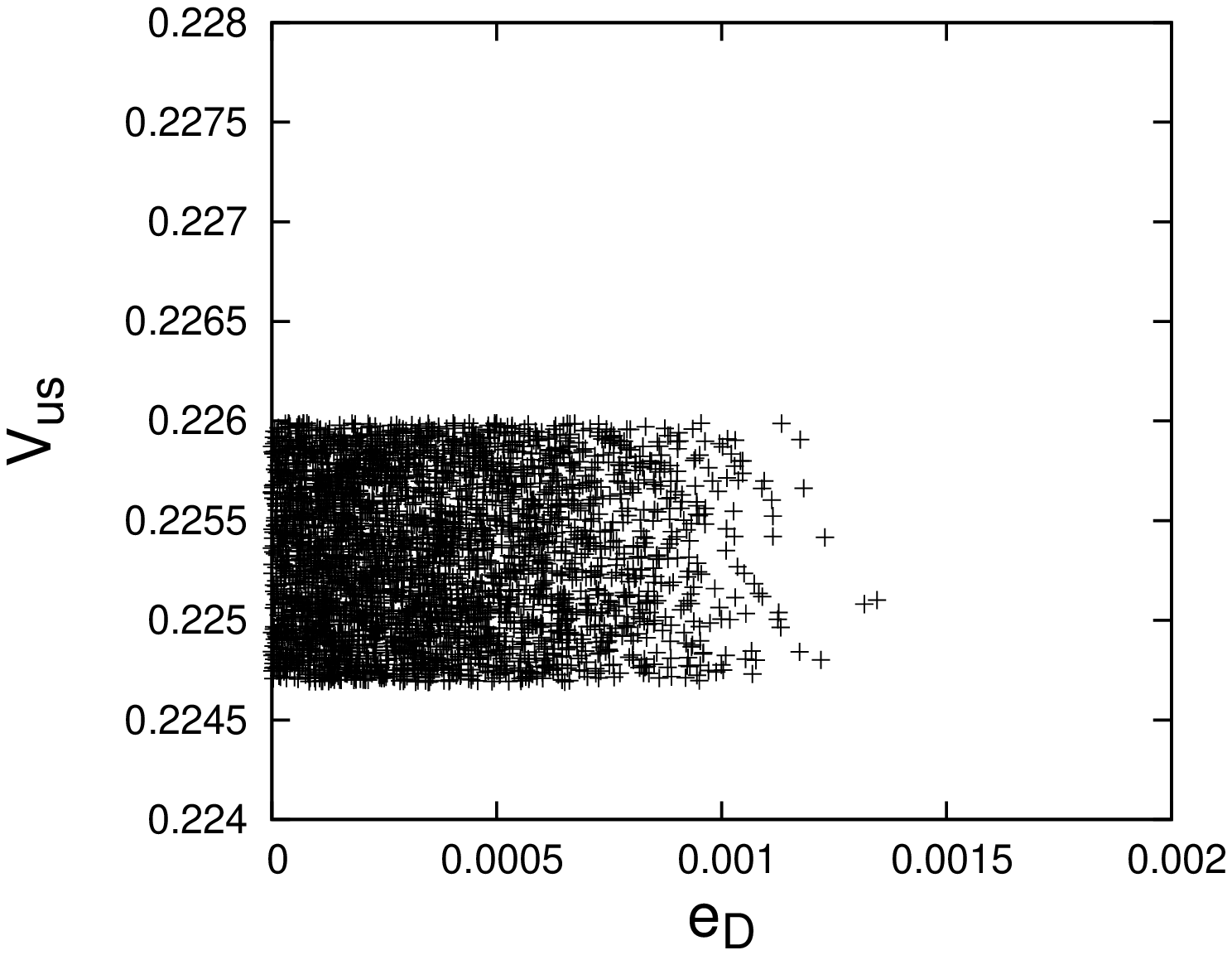}
  \end{minipage}
 \caption{Plots showing variation of
the magnitude of $V_{us}$ with the parameters $e_U$~ and
$e_D$~ for the texture two zero quark mass matrices. }
\label{fig2}
\end{figure}
Interestingly, we find that the elements $e_U$ and $e_D$ do not seem to be
playing any significant role in reproducing the CKM matrix. This can be understood by
considering the effect of variation of $e_U$ and $e_D$ on some of the CKM elements. To this end, in
figure (\ref{fig2}), we have plotted $|V_{us}|$ versus $e_U$ and $e_D$. As is evident
from the figure, $|V_{us}|$ does not show any dependence with the variation of $e_U$ and $e_D$
indicating the redundancy of these elements. This can also be checked further by considering the
variation of other CKM elements with respect to $e_U$ and $e_D$ . Therefore, it seems that we
do not lose any generality if we ignore the elements $e_U$ and $e_D$ and consider reducing
the mass matrices given in eqn. (\ref{fxingt2}) to
\be
 M_{i(i=U,D)}=\left( \ba{ccc}
0 & a _{i} & 0      \\ a_{i}^{*} & d_{i} &  b_{i}     \\
 0 &     b_{i}^{*}  &  c_{i} \ea \right).
 \label{flt40}
 \ee
Interestingly, this structure is very similar to the original
Fritzsch ansatze for the quark mass matrices wherein the elements
$d_i$ are zero. The matrices given above, in fact, can be
characterized as Fritzsch-like texture four zero quark mass
matrices. The analyses of such matrices have been carried out by
several authors\cite{singreview}, however we briefly present the
results of our present analysis. Using the inputs mentioned above
and imposing the constraints given by PDG 2012 data\cite{pdg2012},
\beqn |V_{us}|=0.22534 \pm 0.00065,
|V_{ub}|=0.00351^{+0.00015}_{-0.00014}\\
|V_{cb}|=0.0412^{+0.0011}_{-0.0005},~sin 2\beta= 0.679 \pm 0.020,
\eeqn we obtain the following CKM matrix
 \be
 V_{ckm}=\left( \ba{ccc}
0.9741-0.9744 & 0.2246-0.2259 & 0.00337-0.00365  \\ 0.2245-0.2258
& 0.9732-0.9736 &  0.0407-0.0422    \\ 0.0071-0.0100 &
0.0396-0.0417 &  0.9990-0.9992 \ea \right).
\label{flckm}
\ee
Interestingly, this matrix appears to be in full agreement with the CKM matrix
given by PDG 2012. Similarly, one can also calculate the angles of the unitarity triangle
and Jarlskog's rephasing invariant parameter J as well. Therefore, it seems that at present
the texture four zero Fritzsch-like quark mass matrices are fully compatible with
the present quark mixing data.
\subsection{Non-Fritzsch-like texture four zero quark mass matrices}
Keeping in mind the WB transformations, which also include permutation symmetries, one can
in fact find other texture four zero structures also which are essentially non Fritzsch-like, e.g.,
\be
a:~~\left ( \ba{ccc} d & a & 0 \\ a^{*}  &  0 &  b \\ 0 & b^*  & c
\ea \right )~~~{\rm{and~ its ~ permutations}}, \ee

\be
b:~~ \left ( \ba{ccc}  0 & a &
 d\\ a^*  & 0  & b \\ d^* & b^*  &
c \ea \right ) ~~~{\rm{and~ its ~ permutations}}, \ee

\be
c:~~\left ( \ba{ccc} a &  0 &  0 \\ 0  & d & b\\  0 & b^* & c \ea
\right ) ~~~{\rm{and~ its ~ permutations}}. \ee It is immediately
clear that the matrices corresponding to category `c' are  not
viable as all the matrices in this class correspond to the
scenario where one of the generations gets decoupled from the
other two. Regarding categories `a' and `b', we can carry out an
analysis similar to the one discussed earlier pertaining to
Fritzsch-like texture four zero mass matrices. The corresponding
CKM matrices for classes `a' and `b' then come out to be
\be
V_{ckm}^a=\left(\ba{ccc} 0.9740-0.9744 & 0.2247-0.2260 & 0.0024-0.0099\\
0.2205-0.2256 & 0.9509-0.9727 & 0.0596-0.2172 \\ 0.0140-0.0445 &
0.0584-0.2127 & 0.9905-1.0000\ea \right),
\label{4bckm}
\ee
\be
V_{ckm}^b=\left(\ba{ccc} 0.9736-0.9744 &  0.2247-0.2260 &
0.0098-0.0331\\ 0.2226-0.2278 & 0.9549-0.9719 & 0.0659-0.1937\\
0.00007-0.0340 & 0.0694-0.1928 & 0.9810-0.9976\ea\right).
\label{5ackm} \ee One finds that the CKM matrices mentioned above
do not agree with the one given by PDG 2012, thereby implying that
the corresponding mass matrices appear to be ruled out.
Interestingly, this leads us to arrive at an important conclusion
that in the case of texture four zero mass matrices, the only
viable mass matrices are the Fritzsch-like ones and their
permutations.
\section{Lepton mass matrices \label{t2l}}
After having found that there is a set of quark mass matrices
which can be obtained from the most general mass matrices within
the SM, it becomes desirable to check whether similar structures
can satisfy the lepton mixing data also. It needs to be emphasized
that since the lepton masses and mixings are very different from
those in the quark sector, it is not necessary to impose the kind
of constraints considered for the quark mass matrices. However,
keeping in mind the issue of unified fermion mass matrices as
advocated by Smirnov \cite{smirnov}, one needs to check whether
similar structure with the same constraints can satisfy the lepton
mixing data also. To this end, we have examined
\cite{diractable,majoranaarxive} texture four zero lepton mass
matrices for Dirac as well as Majorana neutrino case using the
latest lepton mixing data. The details in this regard will be
published elsewhere. However, we find that in case we impose
naturalness criterion, inverted hierarchy of neutrino masses is
ruled out for Dirac as well as Majorana neutrinos whereas normal
hierarchy and degenerate scenario are viable for both.
\section{Summary and Conclusions \label{con}}
To summarize, in view of the recent refinements in the quark as
well as lepton mixing data, the issue of formulation of fermion
mass matrices which incorporate the low energy data in both the
sectors poses to be an important issue of the flavor physics. In
this context, keeping in mind various broad guidelines such as
Weak Basis transformations and `natural mass matrices', we have
made an attempt to carry out an analysis of the texture specific
mass matrices pertaining to quark as well as lepton sectors.
Starting with the most general mass matrices, using WB
transformations, one can consider the mass matrices wherein the
(1,3) and (3,1) elements in the up sector as well as down sector
mass matrices are zero. Further, when these mass matrices are
subjected to naturalness condition \\ \bc $(1,i)\lesssim (2,j)
\lesssim (3,3);~~~~~~~~~~i=1,~2,~3, ~~j=2,~3,$ \ec one finds that
the (1,1) element in both the mass matrices takes very small
values and seems largely redundant as far as its implications on
the CKM elements are concerned. In other words, starting with the
most general mass matrices one essentially obtains texture four
zero mass matrices, out of which only the Fritzsch-like mass
matrices and their permutations seem to be the viable option.
\par A corresponding study in the lepton sector for texture four zero Fritzsch-like mass matrices
in the Dirac as well as Majorana neutrino case has been carried out. Our present analysis indicates that these
matrices are compatible with the normal hierarchy and degenerate scenario of neutrino masses whereas for inverted
hierarchy such matrices are ruled out in case the naturalness conditions are imposed. In conclusion, we can perhaps
say that the texture four zero Fritzsch-like mass matrices provide an almost unique class of viable fermion
mass matrices giving vital clues towards unified textures for model builders.
\section*{Acknowledgments}
M.G. would like to thank the organizers of `International
Conference on Flavor Physics and Mass Generation', NTU Singapore
for providing an opportunity to present this work. S.S. would like
to acknowledge UGC, Govt. of India, for financial support. G.A.
would like to acknowledge DST, Government of India (Grant No:
SR/FTP/PS-017/2012) for financial support. S.S., P.F., G.A.
acknowledge the Chairperson, Department of Physics, P.U., for
providing facilities to work.

\end{document}